# The Curious Case of ASAS J174600-2321.3: an Eclipsing Symbiotic Nova in Outburst?

**Stefan Hümmerich**
*Stiftstr. 4, Braubach, D-56338, Germany; American Association of Variable Star Observers (AAVSO), Cambridge, MA; Bundesdeutsche Arbeitsgemeinschaft für Veränderliche Sterne e.V. (BAV), Berlin, Germany; stefan.huemmerich@gmail.com*

**Sebastián Otero**
*Olazabal 3650-8 C, Buenos Aires, 1430, Argentina; American Association of Variable Star Observers (AAVSO), Cambridge, MA*

**Patrick Tisserand**
*Sorbonne Universités, UPMC Univ Paris 06, UMR 7095, Institut d'Astrophysique de Paris, F-75005 Paris, France; CNRS, UMR 7095, Institut d'Astrophysique de Paris, 98 bis Boulevard Arago, F-75014 Paris, France*

**Klaus Bernhard**
*Kafkaweg 5, Linz, 4030, Austria; American Association of Variable Star Observers (AAVSO), Cambridge, MA; Bundesdeutsche Arbeitsgemeinschaft für Veränderliche Sterne e.V. (BAV), Berlin, Germany*

**Abstract**  The star ASAS J174600-2321.3 was found to exhibit peculiar photometric variability (conspicuous brightening of ~4 magnitudes (V), fast luminosity declines, intrinsic pulsations). It was rejected as an RCB candidate in recent investigations on spectroscopic grounds. We have collected and present all available data from public sky surveys, photometric catalogues, and the literature. From an analysis of these data, we have identified ASAS J174600-2321.3 as a long-period eclipsing binary ($P_{orb}$ = 1,011.5 days). The primary star, which is probably a white dwarf, is currently in outburst and exhibits the spectral characteristics of a reddened, early F-type supergiant; the secondary star is a giant of spectral type late M. We discuss the possible origin of the observed brightening, which is related to the primary component. ASAS J174600-2321.3 is most certainly an eclipsing symbiotic binary—probably a symbiotic nova of GCVS type NC—that is currently in outburst. However, further photometric and spectroscopic data are needed to confirm this.

## 1. Introduction

ASAS J174600-2321.3 = EROS2-cg1131n13463 = 2MASS J17460018-2321163, which is situated in Sagittarius at position (J2000) R.A. $17^h 46^m 00.180^s$ Dec. –23° 21' 16.37" (UCAC4; Zacharias *et al.* 2012), was first published as a variable star in the ASAS Catalog of Variable Stars (ACVS; Pojmański *et al.* 2005), where it was catalogued as a miscellaneous variable (type "MISC"). It was found to exhibit peculiar photometric variability (conspicuous rise in brightness, pulsations). Consequently, it was investigated as a candidate R Coronae Borealis (RCB) star during several recent investigations (Tisserand *et al.* 2008; Tisserand *et al.* 2013) but ultimately rejected on spectroscopic grounds, mainly because of a high abundance of hydrogen. Identifications and coordinates of ASAS J174600-2321.3 are given in Table 1.

We have compiled all available data on this object from various sky surveys, photometric catalogues, and the literature; they are presented in Section 2 and analyzed and interpreted in Section 3. Possible reasons for the rise in magnitude and probable classifications are discussed in Section 4. We conclude in Section 5.

## 2. Observations

### 2.1. Multiband photometric data

Before presenting the photometric data, it is important to point out that there is considerable line-of-sight extinction to ASAS J174600-2321.3, which affects the recorded photometric properties of the star. All dust-reddening estimates in this and the following sections are based on the work of Schlafly and Finkbeiner (2011), who employ the colors of stars with spectra in the Sloan Digital Sky Survey (Adelman-McCarthy *et al.* 2011) and measure reddening as the difference between the measured and predicted colors of a star. Their results prefer an $R_V$ = 3.1 Fitzpatrick (1999) reddening law. Based on the calculations of Schlafly and Finkbeiner (2011), we estimate an interstellar extinction of $A_V \approx 2.4$ mag. and $E(B-V) \approx 0.78$ mag. for the sky area of our interest. The extinction values in different bandpasses, which have been used for the reddening correction in the present paper, were accessed through the NASA/IPAC Infrared Science Archive (http://irsa.ipac.caltech.edu/applications/DUST/). If not indicated otherwise, the light curves illustrated in this and the following sections are based on non-corrected data.

Table 1. Identifications and coordinates of ASAS J174600-2321.3. Positional data were taken from UCAC4.

| Identifiers | R.A. (J2000) h m s | Dec. (J2000) ° ' " | Galactic Longitude (°) | Galactic Latitude (°) |
|---|---|---|---|---|
| ASAS J174600-2321.3 EROS2-cg1131n13463 2MASS J17460018-2321163 USNO-B1.0 0666-0562145 | 17 46 00.180 | –23 21 16.37 | 2.829 | 4.814 |

In order to achieve a long time baseline, we have combined photometric observations from the EROS-2 project (Renault *et al.* 1998), ASAS-3 (Pojmański 2002), and APASS (Henden *et al.* 2012). EROS-2 observations were performed with two wide field cameras behind a dichroic cube splitting the light beam into two broad passbands. The so-called "blue" channel (BE, 420–720 nm) overlapped the V and R standard bands; the "red" channel (RE, 620–920 nm) roughly matched the mean wavelength of the Cousins I band. Data from the EROS-2 project have been transformed to Johnson V and Cousins I using Equation (4) of Tisserand *et al.* (2007):

$$R_{eros} = I_C \qquad B_{eros} = V_J - 0.4(V_J - I_C) \qquad (1)$$

The obtained results are accurate to a precision of 0.1 mag. (Tisserand *et al.* 2007). Unfortunately, problems with the CCD detectors of the "red" camera resulted in missing $R_E$ data after HJD 2452200. Thus, no contemporaneous $R_E$ measurements were available to transform the $B_E$ values after HJD 2452200 using Equation (1). The missing data have been linearly interpolated in such a way that the resulting data forms a continuity with the preceding EROS-2 and succeeding ASAS-3 measurements. We have achieved this by using Equation (2), which has been derived from the conversion of data before HJD 2452200. As the system brightened considerably during the time of the missing $R_E$ measurements, and the brightening is obviously related to the hot primary component, we have also taken into account the resulting color changes and have adapted the equation accordingly. (As primary star, we define the star which is the seat of the observed brightening and which we propose as possible accretor in a symbiotic binary scenario.)

$$\Delta I_C = 0.57 \times \Delta B_E \qquad (2)$$

As can be seen from the resulting light curve, there is very good agreement between the converted EROS-2 and ASAS-3 data. Furthermore, the derived color index from the resulting data, $(V-I_C)_{eros} = 1.9$ mag. at V ≈ 13.0 mag., is consistent with the index derived from APASS photometry at maximum brightness, $(V-I_C)_{apass} = 1.5$ mag. at V ≈ 12.2 mag. We thus feel confident about the applicability of our solution.

The combined light curve of ASAS J174600-2321.3, based mainly on sky survey data from EROS-2, ASAS-3, and APASS, is shown in Figure 1; APASS measurements are presented in Table 2.

During the beginning of EROS-2 coverage, the star's brightness varied slowly around 16.5 mag. (V). A rise in magnitude set in at around HJD 2451240, with the system reaching 15 mag. (V) about 350 days later. After a short plateau and the eclipse at around HJD 2452100, ASAS J174600-2321.3 continued to brighten in V during the rest of EROS-2 and the beginning of ASAS-3 coverage, rising by about three magnitudes in ~500 days. The star reached ~12.2 mag. (V) at about HJD 2452600, around which it has apparently remained up to the present time. Interestingly, the conspicuous brightening is not seen at longer wavelengths; the mean brightness of the corresponding EROS-2 $I_C$ measurements increased by only 0.2 mag., although—as mentioned above—there exists no $I_C$ photometry for the most dramatic part of the brightening between about HJD 2452200 and HJD 2452680 because of the missing $R_E$ measurements after HJD 2452200. In order to derive an approximate amplitude in $I_C$, we have transformed APASS photometry at maximum using Equation (3) of Jester *et al.* (2005; especially their Table 1):

$$R - I = 1.00(r - i) + 0.21 \qquad (3)$$

From this, we derive $I_C$ = 10.6 mag. at maximum, which indicates a total amplitude of ~2 mag. in this band—less than half of the amplitude in V.

2.2. The observed outburst and a historical light curve

As becomes obvious from Figure 1, the system shows complex photometric variability on different timescales. The most striking feature of the light curve is the already mentioned drastic increase in visual brightness related to the primary component. This outburst has been accompanied by a blueward color evolution. Before the rise in brightness (HJD < 2451112), we derive a mean extinction corrected color index of $(V-I_C)_0 \approx 3.0$ mag., which is consistent with a spectral type of ~mid-M (Ducati *et al.* 2001). At the last simultaneous V and $I_C$ measurements at HJD 2452200.542—after the system had already brightened by ~1.5 mag. (V)—the color index became bluer, to $(V-I_C)_0 \approx 1.8$ mag. Finally, at maximum visual light, the derived color index, $(B–V)_0 \approx 0.2$ mag., is in agreement with a late A- / early F-type supergiant (see sections 3.2 and 3.3).

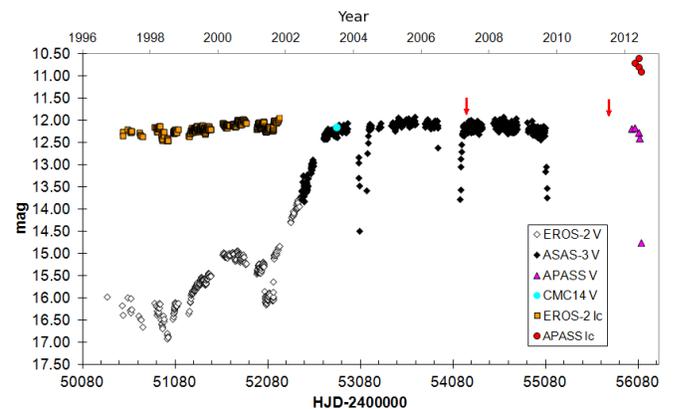

Figure 1. Light curve of ASAS J174600-2321.3, based on EROS-2, ASAS-3, and APASS data. EROS-2 $B_E$ and $R_E$ magnitudes have been transformed to Johnson V and Cousins I, respectively; Cousins I values have been derived from APASS photometry using the equations in Jester *et al.* (2005); see text for details. Also included is a transformed observation from CMC14 (Evans *et al.* 2002). Some obvious outliers with a photometric quality flag of D have been removed from the ASAS-3 dataset. The epochs of the spectra are indicated by arrows; magnitudes were not corrected for extinction (see text for details).

Table 2. Photometric measurements of ASAS J174600-2321.3 from APASS. $I_C$ values have been derived from APASS photometry using the equations in Jester *et al.* (2005).

| Passband | HJD–2400000 | Magnitude | Error |
|---|---|---|---|
| APASS B | 56008.7946 | 13.208 | 0.006 |
| | 56040.7133 | 13.189 | 0.006 |
| | 56085.5371 | 13.347 | 0.010 |
| | 56086.5358 | 13.476 | 0.007 |
| | 56105.6268 | 15.992 | 0.060 |
| APASS V | 56008.7942 | 12.179 | 0.005 |
| | 56040.7128 | 12.176 | 0.005 |
| | 56085.5366 | 12.275 | 0.007 |
| | 56086.5353 | 12.398 | 0.006 |
| | 56105.6263 | 14.753 | 0.042 |
| APASS i' | 56008.7968 | 11.098 | 0.006 |
| | 56040.7154 | 11.274 | 0.008 |
| | 56085.5544 | 11.342 | 0.018 |
| | 56086.5379 | 11.368 | 0.015 |
| | 56101.8428 | 12.445 | 0.026 |
| | 56102.8404 | 12.562 | 0.026 |
| | 56103.8376 | 12.609 | 0.028 |
| | 56105.6939 | 12.502 | 0.029 |
| APASS g' | 56040.7159 | 12.681 | 0.004 |
| | 56085.5422 | 12.817 | 0.006 |
| | 56086.5410 | 12.878 | 0.004 |
| | 56105.6293 | 15.373 | 0.035 |
| | 56134.6033 | 16.700 | 0.109 |
| APASS r' | 56040.7141 | 11.778 | 0.005 |
| | 56085.5379 | 11.861 | 0.006 |
| | 56086.5366 | 12.020 | 0.006 |
| | 56105.6276 | 14.123 | 0.037 |
| | 56134.6016 | 14.689 | 0.060 |
| APASS $I_C$ | 56040.7128 | 10.7 | 0.15 |
| | 56085.5366 | 10.8 | 0.15 |
| | 56086.5353 | 10.6 | 0.15 |
| | 56105.6263 | 10.9 | 0.15 |

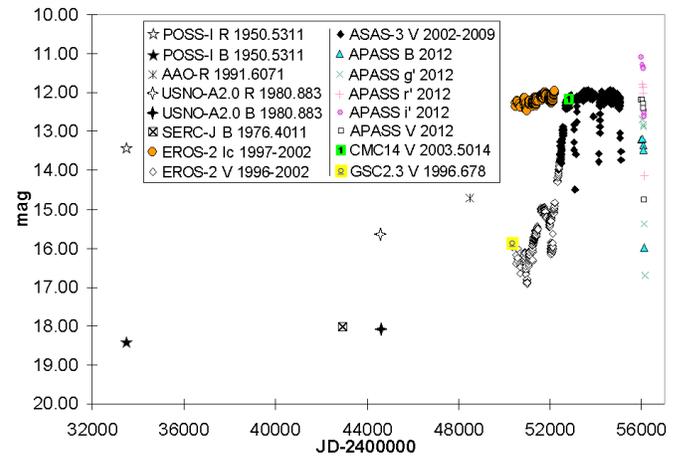

Figure 2. Light curve of ASAS J174600-2321.3, based on data from various historical and contemporary sources, as indicated in the legend. Magnitudes were not corrected for extinction.

Table 3. Measurements of ASAS J174600-2321.3 in various astrometric and photometric catalogues. Magnitudes were not corrected for extinction.

| Value | Source | Epoch | Remarks |
|---|---|---|---|
| 18.38 (B1) | POSS-I | 1950.5311 | (B–R) = 4.95 |
| 13.43 (R1) | POSS-I | 1950.5311 | (B–R) = 4.95 |
| 18.01 (B2) | SERC-J | 1976.4011 | |
| 14.69 (R2) | AAO-R | 1991.6071 | |
| 18.1 (B) | USNO-A2.0 | 1980.883 | (B–R) = 2.8 |
| 15.3 (R) | USNO-A2.0 | 1980.883 | (B–R) = 2.8 |
| 15.86 (V) | GSC2.3 | 1996.678 | |
| 15.1–15.5 (V) | UCAC3 | — | Transformed using 2MASS J–K and APASS r' and V. |
| 12.15 (V) | CMC14 | 2003.5014 | Transformed using APASS r' and V. |

In order to investigate the long-term photometric variability of ASAS J174600-2321.3, we have constructed a light curve including historical data from various catalogue sources, which is shown in Figure 2. Archival plate measurements and other observations over the years indicate that the star's magnitude has been found between magnitudes 17 and 18 on several occasions on blue-sensitive photographic plates (Table 3), which is about 4 to 5 magnitudes fainter than the observed maximum B magnitude of 13.2 from APASS—a value well beyond any possible error in plate reductions. No bright states seem to have been recorded in the past. However, as there are only a few scattered measurements in different passbands before the onset of the EROS project, the evidence is tentative, at best.

2.3. Spectroscopic observations

During a search for new RCB stars in the EROS-2 database, Tisserand *et al.* (2008) identified ASAS J174600-2321.3 as a likely candidate on grounds of its RCB-like photometric variability (fast brightness declines; slow, RCB-like recovery at the beginning of ASAS-3 coverage, intrinsic pulsations). A spectrum of the object was obtained using the Dual-Beam Spectrograph (DBS; Rodgers *et al.* 1988) attached to the 2.3-meter telescope at the Siding Spring Observatory of the Australian National University (ANU). The DBS is a general purpose optical spectrograph, which is permanently mounted at the Nasmyth A focus. The visible waveband is split by a dichroic at around 600 nm and feeds two essentially similar spectrographs, with red and blue optimized detectors. The full slit length is 6.7 arcmin.

The spectrum was taken at maximum visual brightness on HJD ~2454213.9218 (hereafter referred to as the "2007 spectrum") and comprises a wavelength range of 5550 Å < λ < 7500 Å and a 2-pixel resolution of 2 Å. The spectrum is shown in Figure 9.

Using enhanced selection criteria, the star was recovered as an RCB candidate during a more recent search by Tisserand *et al.* (2013). This renewed the interest in the object and, accordingly, it was re-investigated and a spectrum with broad wavelength coverage was obtained. Spectroscopic observations were performed with the Wide Field Spectrograph (WiFeS) instrument (Dopita *et al.* 2007) attached to the ANU 2.3-m telescope at the Siding Spring Observatory. WiFeS is an integral field spectrograph permanently mounted at the Nasmyth A focus. It provides a 25" × 38" field with 0.5-arcsec sampling along each of the twenty-five 38" × 1" slitlets. The visible wavelength interval is divided by a dichroic at around 600 nm feeding two essentially similar spectrographs.

The spectrum, which comprises a 2-pixel resolution of 2 Å, was taken at maximum visual brightness on HJD ~2455762.9661 (the "2012 spectrum"); a segment is shown in Figure 10 (5700 Å < λ < 9700 Å), the entire spectrum is given in Figure 11. The epochs of both spectra are indicated in Figure 1.

## 3. Data analysis and interpretation

We interpret the light curve characteristics (brightening, eclipses, pulsations), the spectral energy distribution, and the spectra as being due to the fact that ASAS J174600-2321.3 is made up of two components. In the following sections, the observed phenomena are discussed in that context.

### 3.1. Orbital period and eclipse characteristics

Three conspicuous fading episodes were covered to varying degrees by ASAS-3; a fourth one is evident from recent APASS data (Figure 1). From their rapid declines, symmetric recoveries, and strict periodicity, we conclude that these events represent deep eclipses with a period of $P_{orb}$ = 1,011.5 days and an eclipse duration of 111 days, which corresponds to ~11% of the period. No trace of a secondary eclipse could be found. A phase plot illustrating the deep eclipses is shown in Figure 3. The following elements have been derived:

$$\text{HJD (Min I)} = 2456142 + 1011.5 \times E \qquad (4)$$

We have searched for additional eclipses of low amplitude in EROS-2 data (i.e., before the most dramatic part of the brightening). There is another eclipse at around HJD 2452100 ("the 2001 eclipse") which exhibits evidence of a pronounced time of totality, during which the primary star is completely hidden by the eclipsing body. It is interesting to note that the pulsational variability—which will be discussed in section 3.4—continues during the observed time of totality (Figure 4).

No other clearly defined eclipses could be found. However, there is a curious, "V-shaped" fading near the expected time of eclipse at around HJD 2451050 ("the 1998 fading"; Figure 5). This is likely not an actual eclipse (it occurs 76.5 ± 25 days before the computed time of minimum as determined by

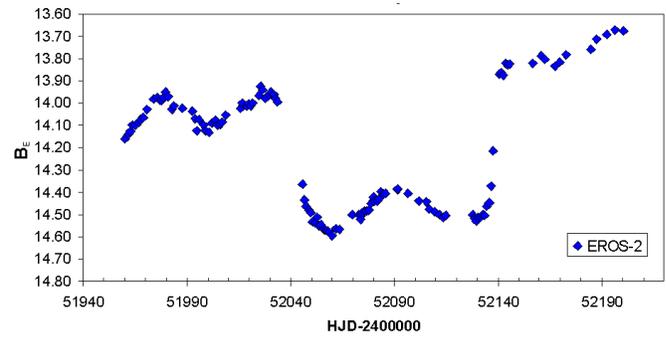

Figure 4. Close-up view of the 2001 eclipse. In order to preserve a maximum number of datapoints, the plot has been based on EROS-2 instrumental $B_E$ magnitudes. Note the continuance of pulsations during the time of totality.

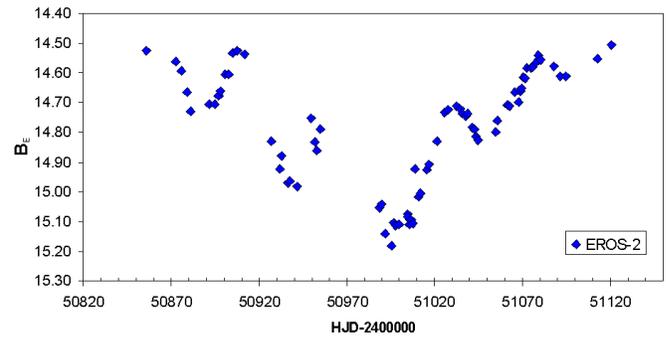

Figure 5. Close-up view of the 1998 fading. In order to preserve a maximum number of datapoints, the plot has been based on EROS-2 instrumental $B_E$ magnitudes.

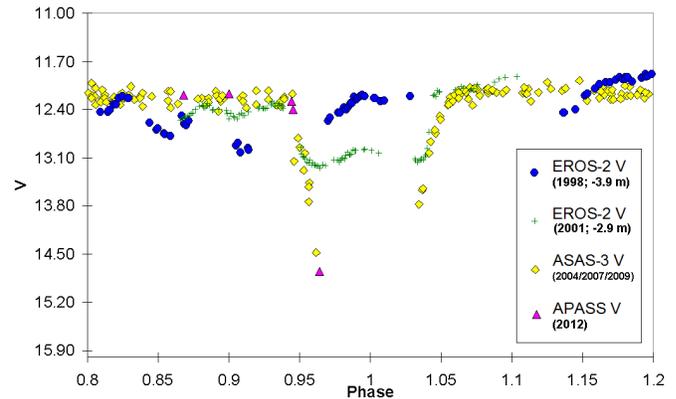

Figure 6. Phase plot of ASAS J174600-2321.3, illustrating all primary eclipses; also included is the 1998 fading episode. The plot has been based on EROS-2, ASAS-3, and APASS V data and folded with the elements given in Equation (4). EROS-2 $B_E$ magnitudes have been transformed to Johnson V (see text for details). In order to facilitate comparison, the resulting EROS-2 V magnitudes have been shifted by the indicated amounts to match V magnitudes from ASAS-3 and APASS.

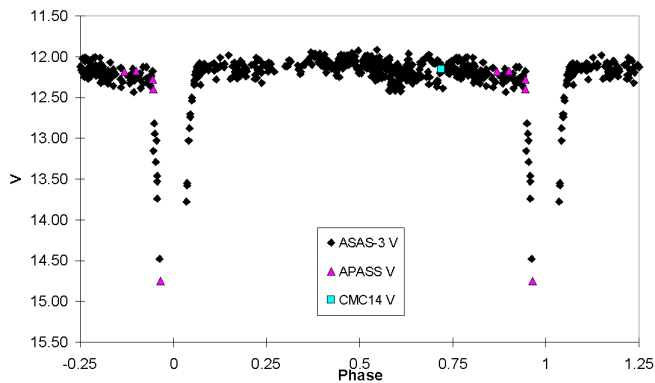

Figure 3. Phase plot of ASAS J174600-2321.3, based on ASAS-3, APASS V, and transformed CMC14 data, and folded with the ephemeris given in Equation (4). Only data > HJD 2452700 have been considered for the phase plot.

Table 4. Recent photometric measurements of ASAS J174600-2321.3. Catalogue data have been accessed through VizieR and corrected for interstellar extinction using the estimates of Schlafly and Finkbeiner (2011), which have been accessed through the NASA/IPAC Infrared Science Archive (http://irsa.ipac.caltech.edu/applications/DUST/). The listed error margins are taken directly from the corresponding catalogues.

| Passband | Source | Magnitude (Error) |
|---|---|---|
| B | APASS (max. vis. light) | 9.99 (±0.02) |
| *B* | *APASS ($\varphi_{orb} \sim 0.964$)* | *12.78 (±0.02)* |
| V | APASS (max. vis. light) | 9.75 (±0.00) |
| *V* | *APASS ($\varphi_{orb} \sim 0.964$)* | *12.33 (±0.00)* |
| g' | APASS (max. vis. light) | 9.69 (±0.01) |
| *g'* | *APASS ($\varphi_{orb} \sim 0.964$)* | *12.39 (±0.01)* |
| r' | APASS (max. vis. light) | 9.71 (±0.01) |
| *r'* | *APASS ($\varphi_{orb} \sim 0.964$)* | *12.06 (±0.01)* |
| i' | APASS (max. vis. light) | 9.65 (±0.13) |
| *i'* | *APASS ($\varphi_{orb} \sim 0.964$)* | *11.07 (±0.13)* |
| J | 2MASS | 9.27 (± 0.022) |
| H | 2MASS | 8.35 (± 0.044) |
| $K_s$ | 2MASS | 8.07 (± 0.024) |
| [3.6] | GLIMPSE | 7.52 (± 0.033) |
| [4.5] | GLIMPSE | 7.60 (± 0.040) |
| [5.8] | GLIMPSE | 7.48 (± 0.039) |
| [8.0] | GLIMPSE | 7.37 (± 0.019) |
| [3.4] | WISE | 7.57 (± 0.026) |
| [4.6] | WISE | 7.62 (± 0.020) |
| [12] | WISE | 7.60 (± 0.022) |
| [22] | WISE | 7.33 (± 0.235) |

| Color Index | Remark |
|---|---|
| $(V–I_C)_0 \approx 3.0$ | mean value at minimum visual light (HJD < 2451112.53) |
| $(B–V)_0 \approx 0.2$ | epoch: HJD 2456008.7942 (maximum visual light) |
| $(B–V)_0 \approx 0.5$ | at $\varphi_{orb} \approx 0.964$ |
| $(J–K_s)_0 \approx 1.2$ | epoch: HJD 2450963.8408 ($\varphi_{orb} \approx 0.88$) (shortly before the rise in mean mag.) |

Equation (4), the uncertainty arising mostly from the pulsations) but might be due to the beating of multiple pulsation modes; however, similar phenomena have been reported in the literature for symbiotic binaries (Skopal 2008, for example). Thus, we feel justified in calling attention to this phenomenon, which will be further discussed in section 4. A phase plot of all eclipses which also includes the 1998 fading is shown in Figure 6.

From the considerable increase in eclipse amplitude after the system's rise in mean magnitude, it becomes obvious that the observed brightening is restricted to the primary star of the system. The contribution of the secondary star, whose light dominates at longer wavelengths, remains practically constant, as can be inferred from the EROS-2 $I_C$ measurements (Figure 1).

Obviously, the contribution of the secondary star to the total flux of the system at visual wavelengths is negligible after the considerable brightening. This is further substantiated by the absence of a secondary minimum and an analysis of recent APASS multi-color photometry, which indicates that the system, as expected, gets redder during eclipses, and the amplitude of the eclipses decreases towards longer wavelengths.

The (B–V) index at maximum light is around 1.0 magnitude, which—after correcting for line-of-sight extinction (section 2.1)—corresponds to $(B–V)_0 \approx 0.2$ mag., getting progressively redder as eclipse sets in. Although there are no observations at mid-eclipse, APASS measurements at a phase of $\varphi \approx 0.964$, which is equivalent to 36.414 days before mid-eclipse, indicate a B–V index of 1.24 mag. (approximating to $(B–V)_0 \approx 0.5$ mag.) and a decrease of 2.6 mag. in V and 1.15 mag. in the Sloan i' band relative to the maximum magnitude. Additional observations at phase $\varphi \approx 0.993$ (7.081 days before mid-eclipse) show the star below the detection limit in the B and V bands while the observed decrease in brightness relative to the measurements at phase $\varphi \approx 0.964$ are only 1.3 mag. and 0.57 mag. in the g' and r' bands, respectively, which confirms the shrinking of the amplitude towards the red end of the spectrum. This is illustrated in Figure 7, which shows a detailed view of the primary eclipse based on BVg'r'i' photometry from APASS and V data from ASAS-3. These findings are also confirmed by EROS-2 data; the amplitude of the 2001 eclipse is 0.2 mag. ($I_C$) and 0.9 mag. (V), respectively.

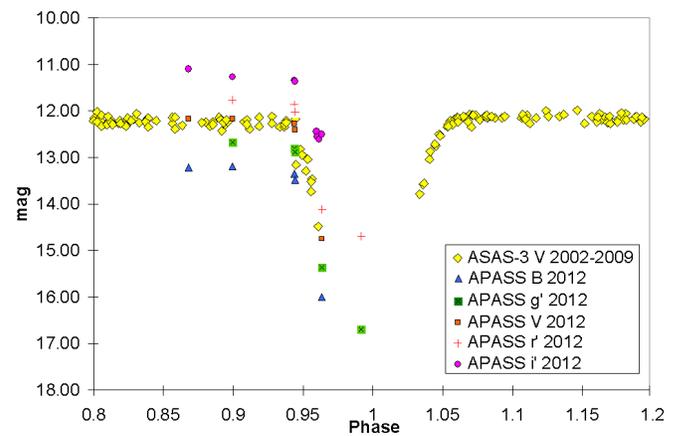

Figure 7. Close-up view of the primary eclipse. The plot is based on ASAS-3 V and APASS multi-color photometry, as indicated in the legend on the right side. The data have been folded with the ephemeris given in Equation (4). Note the decreasing amplitude of the eclipse towards longer wavelengths.

3.2. Spectral energy distribution

Recent photometric measurements from various online catalogue sources are presented in Table 4. Because of the significant line-of-sight dust extinction to ASAS J174600-2321.3 ($A_V \approx 2.4$ mag. and $E(B–V) \approx 0.78$ mag.; Schlafly and Finkbeiner (2011); section 2.1), extinction-corrected values are given for the wavelength range 0.43 µm (Johnson B) < λ < 7.68 µm (GLIMPSE [8.0]).

The spectral energy distribution (SED) of ASAS J174600-2321.3, based on the extinction-corrected values, is shown in Figure 8. APASS data in all five passbands were available from two epochs; APASS data taken at maximum visual light are denoted by triangles, APASS data taken near mid-eclipse at an orbital phase of $\varphi \approx 0.964$ (italicized in Table 4), which corresponds to 36.414 days before mid-eclipse, are denoted by squares.

According to our expectations, the hot component, which—anticipating our results—is likely a white dwarf (WD), dominates the system's emission at maximum visual light, contributing strongly to the short wavelength region and being thus responsible for the observed rise in visual brightness. The red giant, on the other hand, dominates at longer wavelengths, peaking in flux at about 1.65 µm. Near mid-eclipse (orbital phase φ ≈ 0.964), however, the continuum bluewards of λ ~ 1 μm drops off dramatically as the WD is eclipsed by the red giant star, which dominates the system's optical brightness during this part of the orbital cycle. This confirms the simultaneous presence of two different temperature regimes in ASAS J174600-2321.3, which is typical for symbiotic stars (Allen 1984, for example).

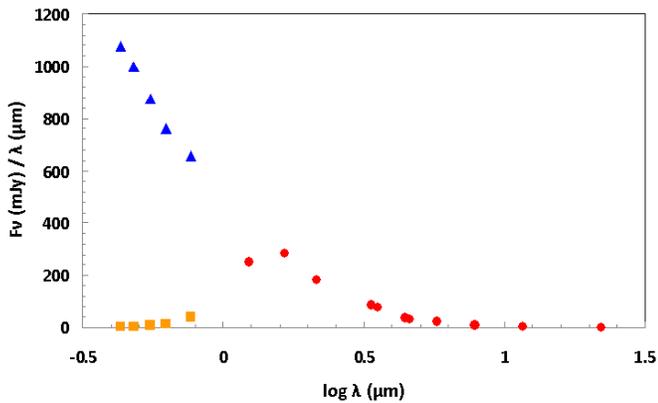

Figure 8. Spectral energy distribution (SED) of ASAS J174600-2321.3 from optical (Johnson B) to mid-infrared (WISE [22]). APASS data taken at maximum visual light are denoted by triangles, APASS data taken near mid-eclipse at an orbital phase of φ ≈ 0.964 are denoted by squares.

3.3. Spectroscopic characteristics and the types of the primary and secondary stars

The most striking features of the 2007 spectrum (shown in Figure 9) are the sodium (Na I) doublet lines at 5889 Å and 5895 Å, and H-alpha (Hα) absorption at 6563 Å. There is also structure due to titanium oxide (TiO) around 6300 Å, 7200 Å, as well as vanadium oxide (VO) absorption at around 7450 Å. The spectrum in this wavelength region resembles that of an M-type giant; consequently, ASAS J174600-2321.3 was classified as spectral type M0 and rejected as an RCB candidate by Tisserand *et al.* (2008).

The increased wavelength coverage (3400 Å < λ < 9600 Å) of the 2012 spectrum (shown in Figures 10 and 11), which includes the Balmer and Paschen lines of hydrogen, shed new light on the star. Based on the 2012 spectrum, which is in agreement with the general characteristics of the 2007 spectrum, the star was re-classified as a highly reddened supergiant of spectral type F0 (Tisserand *et al.* 2013). Reinvestigating the spectrum, we confirm this classification, which is in agreement with the de-reddened color index at maximum visual light of $(B–V)_0 ≈ 0.2$ mag. (section 2.2). It is interesting to note that, in contrast to the 2007 spectrum, the 2012 spectrum displays Hα in emission.

Additionally, as in the 2007 spectrum, there is clearly structure due to TiO absorption around 6300 Å, 7200 Å, 7800 Å, and 8300 Å; VO features are present around 7450 Å and, possibly, 7900 Å (Figure 10). The spectrum is obviously composite, the observed TiO and VO absorption being due to the presence of the cool giant, which agrees well with the observed binarity of the system. Judging from the presence of VO bands, which set in at around M7 and increase in strength with declining temperature (Gray and Corbally 2009, for example), the observed $(V–I_C)_0$ index at minimum light (section 2.2), and the SED (section 3.2), we conclude that the secondary component of ASAS J174600-2321.3 is a late-M giant (~M7), which perfectly fits all observed parameters.

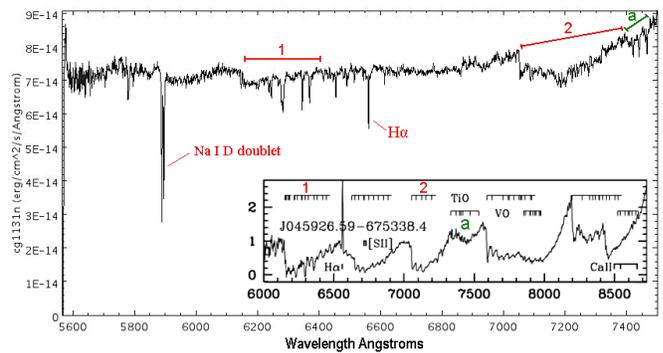

Figure 9. The 2007 spectrum of ASAS J174600-2321.3 (Tisserand *et al.* 2008). The spectrum was taken outside of an eclipse on HJD ~2454213.9218 and encompasses a wavelength range of 5550 Å < λ < 7500 Å. Note the structure due to TiO (identified by numbers) and VO (identified by letters). In order to facilitate discrimination, the spectrum of a late M-type star (2MASS J04592660–6753383) is shown for comparison (inset), which has been adapted from Wood *et al.* (2013; ©2013 The Authors. By permission of Oxford University Press on behalf of the Royal Astronomical Society. All rights reserved. For permissions, please email: journals.permissions@oup.com).

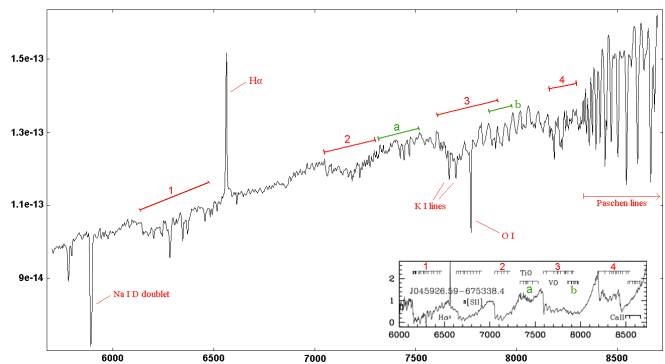

Figure 10. Segment of the 2012 spectrum of ASAS J174600-2321.3 (5700 Å < λ < 8700 Å; Tisserand *et al.* 2013). The spectrum was taken outside of an eclipse on HJD ~2455762.9661. Note the structure due to TiO (identified by numbers) and VO (identified by letters) and the change from absorption (2007 spectrum) to emission (2012 spectrum) in the Hα line at 6563 Å. In order to facilitate discrimination, the spectrum of a late M-type star (2MASS J04592660-6753383) is shown for comparison (inset), which has been adapted from Wood *et al.* (2013; ©2013 The Authors. By permission of Oxford University Press on behalf of the Royal Astronomical Society. All rights reserved. For permissions, please email: journals.permissions@oup.com.). Note the CCD fringing that affects the reddest part of the spectrum. The Paschen lines, however, emerge well above the fringing level for a confident identification.

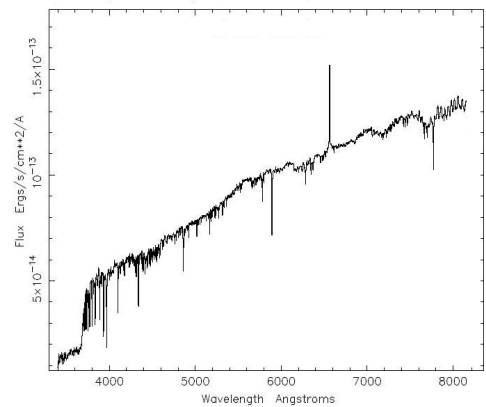

Figure 11. The 2012 spectrum of ASAS J174600-2321.3 to about 8200 Å. The spectrum was taken by Tisserand *et al.* (2013) outside of an eclipse on HJD ~2455762.9661 and encompasses a wavelength range of 3400 Å < λ < 9600 Å.

### 3.4. Pulsation study

In addition to the mean brightness changes, ASAS J174600-2321.3 also shows semiregular pulsations, which are most obvious in EROS-2 $I_C$ data, for example in and around the 2001 eclipse at around HJD 2452100 (Figure 12). It is obvious that the pulsations continue during the total phase of the observed eclipse, which is proof that the pulsations arise in the late M-type secondary component, as would be expected. Although rather ill-defined throughout most of the covered timespan, there is possible pulsational activity with a timescale around 50 to 60 days in EROS-2 data.

After the significant rise in mean brightness related to the primary star, the oscillations are much less clear but still readily visible during maximum light in the ASAS-3 V light curve. However, a period of about 80 to 90 days is now prevalent, especially around HJD 2453500 (Figure 13). While there is no doubt that the pulsations arose in the secondary star before the observed brightening, it is open to debate where they stem from after this event. As the expected contribution of the secondary star at visual wavelengths will now be negligible, the observed pulsations after the brightening are likely to arise in the F-type primary star, which might explain the apparent change in period; however, more data are needed to resolve this matter.

As will be shown in the discussions in section 4, the primary star is likely a WD currently in outburst. In this respect, it is interesting to mention that—according to theoretical considerations—the outer portion of the white dwarf envelope might become pulsationally unstable during the first few months after visual maximum has been reached in a symbiotic nova eruption (Kenyon and Truran 1983). Pulsations of comparable period and amplitude have, for example, been observed during the maximum of the symbiotic nova PU Vul (P ≈ 80 d; ΔV ≈ 0.15 mag.; see Kenyon 1986 and references therein).

An analysis of EROS-2 $I_C$ data with different period search algorithms (CLEANest (Foster 1995) and ANOVA (Schwarzenberg-Czerny 1996), as implemented in PERANSO (Vanmunster 2007); PERIOD04 (Lenz and Breger 2005)) suggests a dominant signal at P ≈ 56 days (Figure 14). Although the amplitude of the signal is rather weak and the error margin is considerable due to the semiregular nature of the variability, we adopt this value as a starting point for the computation of distance and other stellar parameters in the following section.

### 3.5. Distances
#### 3.5.1. Distance in a high extinction area

We have estimated the distance to ASAS J174600-2321.3 using the period-luminosity (P-L) relations for M-type semiregular variables based on Hipparcos parallaxes proposed by Yeşilyaprak and Aslan (2004). Assuming that the M giant dominates the system's light before the outburst in V and, especially, in $I_C$, we have derived mean magnitudes of ASAS J174600-2321.3 in both passbands from converted EROS-2 observations before the onset of the brightening at around HJD 2451240 (V = 16.3 mag.; $I_C$ = 12.2 mag.). We have used the corresponding P-L relations for the Johnson V and Cousins I bands (Yeşilyaprak and Aslan 2004; especially their Table 2), adopting a period value of P = 56 days (section 3.4):

$$M_V = 2.89\,(\pm 0.49) \times \log(P) - 5.30\,(\pm 0.85) \quad (5)$$

$$M_{Ic} = 0.83\,(\pm 0.48) \times \log(P) - 4.79\,(\pm 0.85) \quad (6)$$

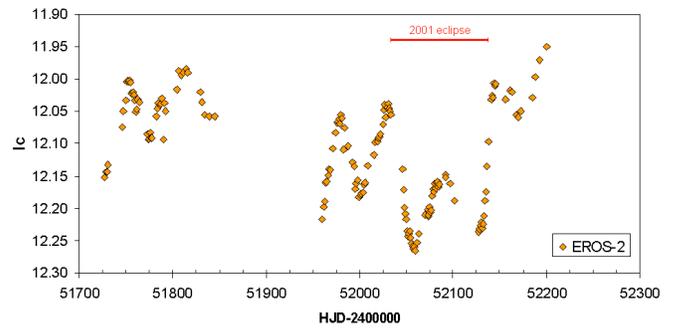

Figure 12. Semiregular pulsations of ASAS J174600-2321.3, based on EROS-2 data (2451726 < HJD < 2452200). EROS-2 $R_E$ magnitudes have been transformed to Cousins $I_C$ (see text for details). Note the continuance of the pulsations during the 2001 eclipse at around HJD 2452100.

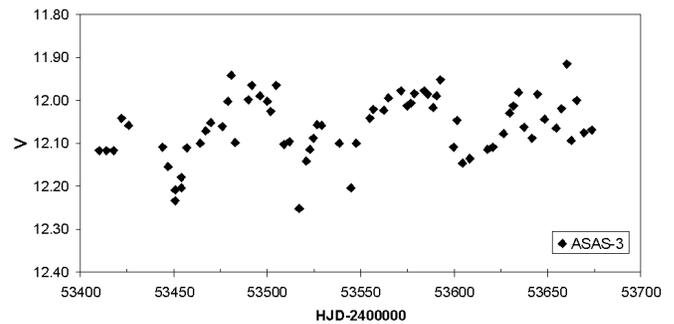

Figure 13. Semiregular pulsations of ASAS J174600-2321.3 after the dramatic brightening, based on ASAS-3 data (2453409 < HJD < 2453673).

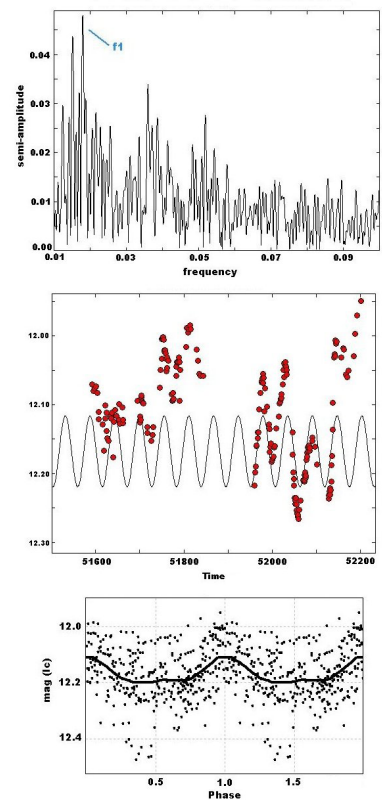

Figure 14. PERIOD04 Fourier graph and period fit (upper panels), based on an analysis of EROS-2 $I_C$ data. The corresponding phase plot (lower panel, with fit curve) is folded with a period of P = 56 days. EROS-2 $R_E$ magnitudes have been transformed to Cousins I (see section 2.1 for details). Note the changes in mean magnitude which are evident from the period fit and phase plot.

From this, we derive absolute magnitudes of $M_V \approx -0.25$ mag. and $M_{Ic} \approx -3.35$ mag. for the red giant component of ASAS J174600-2321.3. Taking into account the line-of-sight extinction $A_V \approx 2.4$ mag. and $A_I \approx 1.4$ mag., calculated using the estimates of Schlafly and Finkbeiner (2011), we derive distances of ~6.7 kpc and ~6.8 kpc from Equations (5) and (6), respectively. Similar results are obtained using the Hipparcos K-band P-L diagram for semiregular variables from Bedding and Zijlstra (1998), which further substantiates our estimates.

Considering the sources of error (intrinsic uncertainties of the P-L relations, conversion of EROS data to standard passbands, semiregularity of the period), it becomes obvious that our distance estimate is only a first approximation that further studies may build upon in the future. However, we feel justified in concluding that ASAS J174600-2321.3 (Galactic coordinates l, b = 2.829°, 4.814°) is situated in the Galactic Bulge, at a distance of about 2 kpc from the Galactic Center.

3.5.2. M giant radius, mass estimates, and orbital distance

We have assessed the effective temperature of the red giant from the proposed spectral class of ~M7 (section 3.3) and derive $T_{eff} \approx 3{,}150$ K. From period and amplitude of the observed pulsations, we have assessed the luminosity of the giant at $\log(L_\star / L_\odot) \approx 3.64$ using Figure 6.2 of Fraser (2008).

By comparing these results to the evolutionary tracks for AGB stars of about solar metallicity (Marigo *et al.* 2008 and Girardi *et al.* 2010; http://stev.oapd.inaf.it/cgi-bin/cmd), we derive a mass of $M_g \approx 1.5\,M_\odot$ for the red giant star (Figure 15), which is in good agreement with the mass estimates for symbiotic binaries compiled from the literature by Mikołajewska (2003, 2010). Based on the aformentioned sources, we furthermore adopt a mass of $M_{wd} \approx 0.5\,M_\odot$ for the white dwarf.

Assuming that the red giant star pulsates in the first overtone, as is observed and proposed for many semiregular variables of similar period (for example, Percy and Parkes 1998; Soszyński *et al.* 2013), we have estimated the radius of the red giant using Equation (4) of Mondal and Chandrasekhar (2005):

$$\log P = 1.59 \log(R/R_\odot) - 0.51 \log(M/M_\odot) - 1.60 \quad (7)$$

Using P = 56 days and $M_g = 1.5\,M_\odot$, we derive a radius of $R_g \approx 145\,R_\odot$, which is in accordance with the values gleaned from the "Catalogue of Apparent Diameters and Absolute Radii of Stars" (Pasinetti Fracassini *et al.* 2001) for other stars of similar spectral type and period.

Using the third Keplerian law and assuming a circular orbit, we have estimated the distance between the red giant and the WD to be $R \approx 2.5$ AU. This results in an essentially consistent picture that is in general accordance with the data compiled by Mikołajewska (2003) for other systems. Furthermore, the adopted configuration is able to reproduce the observed eclipse duration of about 111 days (section 3.1).

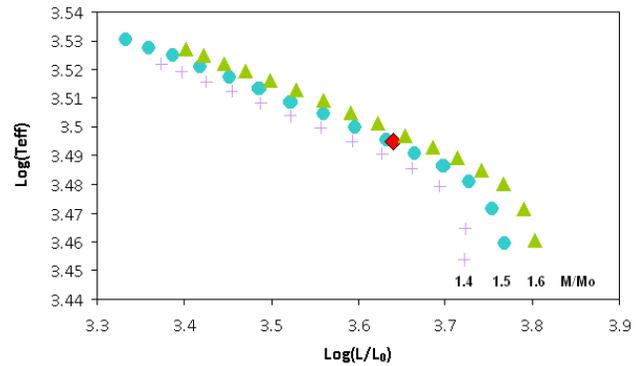

Figure 15. Stellar evolutionary tracks for AGB stars of about solar metallicity and different masses, adapted from Marigo *et al.* (2008) and Girardi *et al.* (2010) and accessed through http://stev.oapd.inaf.it/cgi-bin/cmd. The position of the red giant component of ASAS J174600-2321.3 is marked by the diamond.

We used a simplified model, therefore the given values have to be treated with caution and serve only as a first approximation. For example, it is possible that the radius of the red giant star has been underestimated because of the unknown exact orbital inclination of the system. Only detailed studies will be able to shed more light on the system's parameters.

**4. Discussion**

4.1. General discussion

Taking into account all evidence, we conclude that ASAS J174600-2321.3 consists of a primary star that exhibits the spectral features of an early F-type supergiant in outburst and a secondary component of spectral type late M (~M7). The primary star, as will be shown in the following discussions, is a hot, compact object – probably a WD. Considering the orbital period, the observed time of totality, its significant contribution to the total flux of the system at longer wavelengths, and its intrinsic variability, we deduce that the cool component is an evolved giant.

Judging from the light curve peculiarities and the observed spectroscopic features (notably the Hα emission in the 2012 spectrum), it seems likely that both components interact. Circumstellar (or circumbinary) gas and dust might be involved, either in a disk or shell. Following this scenario, the observed long-term rise in brightness might be due to the dispersion of an optically thick circumstellar shell of gas and dust—a phenomenon regularly seen in young or evolved stars of high luminosity. However, this is not supported by near- and mid-infrared measurements. Furthermore, no bright state has been recorded during the past, although this evidence is based on only a few scattered measurements before the onset of the EROS project (section 2.2). Nevertheless, it seems likely that the observed brightening does not constitute the recovery from a faint state but rather an outburst. Obviously, then, being the seat of the brightening, the primary star has developed the observed F-type supergiant features in outburst.

Symbiotic stars (GCVS-type ZAND; Samus *et al.* 2007–2013) and symbiotic novae (also called "very slow novae"; GCVS-type NC) are subject to prolonged outbursts and associated with long-period binary systems. The symbiotic star/nova scenarios are explored in more detail in the following sections.

4.2. The symbiotic star scenario

The term symbiotic star was coined by Merril in 1941 and applied to stars whose spectra are characterized by a combination of the features of a hot source and a cool continuum with prominent absorption features reminiscent of a late-type star (see for example, Sahade 1982; Crocker *et al.* 2001). In fact, symbiotic stars are long-period interacting binary systems, comprising a compact, hot object as accretor (usually a white dwarf), and a yellow or red giant star as donor (for example, Kenyon 1986; Skopal 1998). These systems are characterized by complex photometric variability, combining eclipses, pulsations, and outbursts on diverse scales, which is akin to what is seen in ASAS J174600-2321.3.

As has been pointed out above, ASAS J174600-2321.3 exhibits a composite spectrum characterized by the features of an early F-type supergiant, which might originate from the pseudophotosphere of a WD in eruption, and the molecular absorption lines of a late M star. It also exhibits photometric variability characteristic of symbiotic stars, which will be discussed below. Several characteristic light curve features are indicated in Figure 16 and marked by the corresponding lower case letters used in the explanatory text below.

(a) During quiet phases, the light curves of symbiotic stars are characterized by wave-like orbitally related variations (for example, Skopal 2008; Skopal *et al.* 2012; their Figures 15 and 16). During quiescence, radiation from the hot component (partly) ionizes the wind from the late-type giant and the neutral circumbinary material, giving rise to a local $H_{II}$ region, whose nebular radiation—according to Skopal (2008)—is the source of the orbitally related variations. Other phenomena, like ellipsoidal variations, might be involved (Gromadzki *et al.* 2013). In amplitude and shape, those variations are reminiscent of the "arc" seen in the light curve of the present object between HJD ~2451200 and HJD ~2452100, before the onset of activity in the system.

(b) In symbiotic systems with high orbital inclination, significant changes in the minima profiles are observed between quiescence and activity. The local $H_{II}$ region has an asymmetrical shape due to orbital motion and interaction with the stellar winds. The nebular emission varies as a function of orbital phase, producing broad, flat minima during quiescence, which occur prior to the time of spectroscopic conjunction at an orbital phase of $\varphi \approx 0.9$ (for a more detailed discussion, see Skopal 1998 and Skopal 2008). This leads to systematic changes in the O–C residuals of the minima times in eclipsing symbiotic binaries (for example, Skopal 2008, especially section 6). As has been described in section 3.1, the "V-shaped" 1998 fading occurred at an orbital phase of $\varphi \approx 0.9$. It is comparable in amplitude and shape to what has been described for other symbiotic binaries (Skopal 2008). It is thus intriguing to interpret the 1998 fading in this vein; however, as has been pointed out above, it may simply be the result of the interference of different pulsation modes. Only long-term light curve studies will be able to shed more light on this issue.

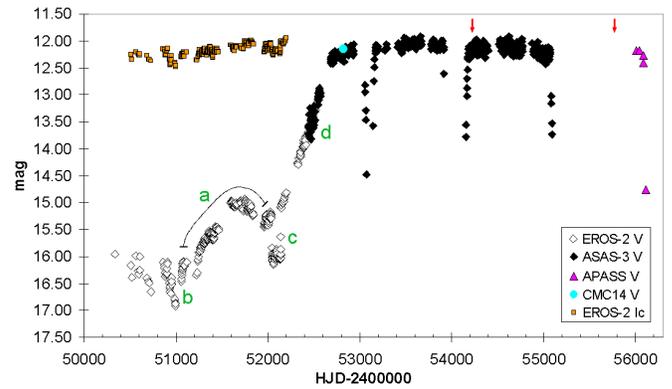

Figure 16. Light curve of ASAS J174600-2321.3, based on EROS-2, ASAS-3 and APASS V data. Symbols and data sources are the same as in Figure 1. The epochs of the spectra are indicated by arrows (see section 3.3 for details). Several light curve features, discussed in the text, are marked by lowercase letters.

(c) With the onset of activity in the system, the optical region usually becomes dominated by the radiation from the pseudophotosphere of the compact object and a narrow, "Algol-like" eclipse is observed at the inferior conjunction of the giant (Skopal 2008). This is exactly what has been observed in the present object after the onset of activity that led to the dramatic brightening of the primary component.

(d) Several types of brightenings are observed in symbiotic binaries (for example, Leibowitz and Formiggini 2006; Skopal *et al.* 2012), encompassing slow, symbiotic-nova-type outbursts, short-term flares, and ZAND-type outbursts of medium amplitude (~3 mag. in V). Active phases can last for many years, which also seems to be the case with ASAS J174600-2321.3. However, the observed outburst differs from that of classical symbiotic stars. The amplitude is larger, and the hot star comes to completely dominate the optical emission; superposed irregular variations are not seen and the outburst is very slow and much quieter than in other sources, which is reminiscent of the behavior of symbiotic novae.

4.3. An eclipsing, symbiotic nova?

Symbiotic novae are a small class of objects that have been variously classified as either classical novae (for example, RT Ser, RR Tel) or symbiotic stars (for example, V1016 Cyg, HM Sge) in the literature (Kenyon and Truran 1983). They are a rare subclass of symbiotic stars (e.g. Kenyon 1986) and distinguished from the classical specimens by their outburst activity (occasional eruptions on timescales from months to years and amplitudes from 1 to 3 mag. for ZAND systems; a single outburst of up to ~7 mag. lasting for decades for NC systems; see, for example, Kenyon and Truran 1983).

Symbiotic novae are themselves often subdivided into two groups: very slow novae (GCVS-type NC) or very fast recurrent novae (GCVS-type NR) with recurrence times of the order of several years (Kato 2002), the deciding criterion being the mass of the WD. According to the aforementioned author, the term symbiotic nova is usually reserved for the former class of objects, which we adhere to throughout this paper; the latter group is of no interest in the analysis of the present object.

Symbiotic novae eruptions develop very slowly; the systems may remain at maximum light for decades, after which they fade very slowly. Apart from that, symbiotic novae share common properties with classical symbiotic stars: they are always long-period ($P_{orb} \geq 800$ days; Mikołajewska 2010), interacting binary stars comprising an M giant. In fact, there exists no clear cut boundary between classical symbiotic stars and symbiotic novae (Skopal *et al.* 2012, for example).

As the underlying mechanism in symbiotic novae eruptions, Kenyon and Truran (1983) propose hydrogen shell flashes involving low-luminosity WDs accreting matter at low rates. Flashes in a degenerate WD envelope ("degenerate flash") lead to evolution from a relatively faint object to a luminous A-F supergiant at maximum visual light, which is in agreement with the spectroscopic characteristics of the present object observed at maximum visual brightness. Systems with WDs undergoing degenerate flashes (as is proposed for RR Tel and RT Ser) remain at visual maximum for decades and lose mass in a low velocity wind (Kenyon and Truran 1983). The higher excitation emission lines, characteristic of most symbiotic systems, fade during the rise in visual brightness (Kenyon and Truran 1983; Corradi *et al.* 2010) and return during the decline, indicating that the WD increases again in effective temperature.

The above mentioned findings are in agreement with the system configuration and spectral characteristics of ASAS J174600-2321.3. No high excitation emission lines are present in our spectra and there is no clear evidence for a rapidly expanding envelope, except for the change in the Hα line from absorption in the 2007 spectrum to emission in the 2012 spectrum, which might be interpreted along the lines of mass loss in a low velocity wind.

Furthermore, as has been pointed out in section 4.2, the general characteristics of the observed rise in brightness in the present object are similar in respect to amplitude and maximum peak shape to what has been observed for many symbiotic novae, see, for example, the historical light curve of RR Tel shown in Figure 17 or the optical light curve of V1016 Cyg in Kenyon (1986; Figure A.11). According to ASAS-3 and APASS data, the system has remained at visual maximum light for more than ~3,200 days. Furthermore, the star is completely saturated in OGLE-III and OGLE-IV data, which reaches up to the present time (Soszyński, 2014; private communication). Additionally, we have started a long-term photometric monitoring program on the star, the results of which will be presented in an upcoming paper. First results haven shown that, in the timespan from HJD 2456855.6 to HJD 2456883.5, the star's brightness fluctuated around a mean magnitude of 12.27 mag (V), which is strong evidence that the outburst has already continued for more than ~4,100 days and still continues.

In summary, an accretion scenario involving the hot primary star of ASAS J174600-2321.3 as accretor and the semiregular, late-type secondary component as donor seems to be the most promising scenario which is capable of explaining all observed light curve features. Thus, we feel confident that the star is an eclipsing symbiotic system. Because of the peculiarities of the observed outburst (amplitude, duration, shape), we propose ASAS J174600-2321.3 as a symbiotic nova of type NC. However, further photometric and spectroscopic data are needed to confirm this.

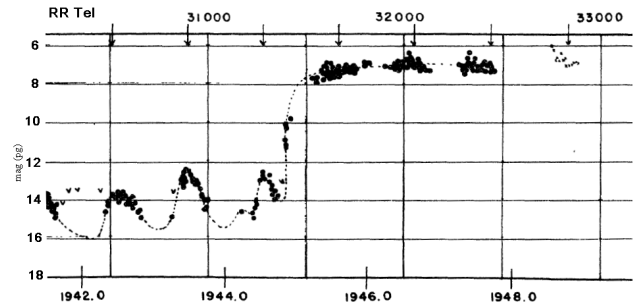

Figure 17. Historical light curve of RR Tel, adapted from Mayall (1949; data prior to 1948 are photographic from Harvard College Observatory plates; 1948 data are visual observations from the AAVSO).

## 5. Conclusion

By an analysis of all available data from the literature, public sky surveys, and photometric catalogues, we have identified the star ASAS J174600-2321.3 as a long-period eclipsing binary with an orbital period of $P_{orb} = 1,011.5$ days. The primary star, which is likely a white dwarf in outburst, exhibits spectral characteristics of a reddened, early F-type supergiant at visual maximum light; the secondary star is a giant of spectral type late M (~M7). The system underwent a significant increase in brightness (~4 mag (V)) during the recent past, which is related to the primary component. We have discussed the origin of the observed brightening and favor the scenario of an eclipsing, symbiotic binary, which is capable of explaining the observed, complex light curve features. Because of the peculiarities of the observed outburst (large amplitude, duration, slow development) and spectral characteristics, we propose ASAS J174600-2321.3 as a symbiotic nova of type NC. Magnitude range, periods, stellar parameters, and binary separation of ASAS J174600-2321.3, as derived in the present paper, are presented in Table 5.

The outburst has already lasted for more than ~4,100 days and seems to continue to the present day. We have started a long-term photometric monitoring of the system, the results of which will be presented in an upcoming paper.

## 6. Acknowledgements

This research has made use of the SIMBAD and VizieR databases operated at the Centre de Données Astronomiques (Strasbourg) in France. This work has also made use of EROS-2 data, which were kindly provided by the EROS collaboration. The EROS (Expérience pour la Recherche d'Objets Sombres) project was funded by the CEA and the IN2P3 and INSU CNRS institutes. Furthermore, this research has employed data products from the Two Micron All Sky Survey, which is a joint project of the University of Massachusetts and the Infrared Processing and Analysis Center/California Institute of Technology, funded by the National Aeronautics and Space Administration and the National Science Foundation, and the

Table 5. Magnitude range, periods, stellar parameters, and binary separation of ASAS J174600-2321.3, as derived in the present paper.

| Parameter | Value |
|---|---|
| Magnitude Range (V) | 11.9–16.9 mag. |
| Orbital Period | 1,011.5 days |
| Stellar Parameters | |
| (Red Giant) | $M_g \approx 1.5\ M_\odot$ |
| | $R_g \approx 145\ R_\odot$ |
| | $T_{eff} \approx 3{,}130$ K |
| | Spectral type ~M7 |
| (White Dwarf) | $M_{wd} \approx 0.5\ M_\odot$ |
| Binary Separation | $R \approx 2.5$ AU |

Wide-field Infrared Survey Explorer, which is a joint project of the University of California, Los Angeles, and the Jet Propulsion Laboratory/California Institute of Technology, funded by the National Aeronautics and Space Administration. This research has also made use of the NASA/IPAC Infrared Science Archive, which is operated by the Jet Propulsion Laboratory/California Institute of Technology, under contract with the National Aeronautics and Space Administration. The authors would like to thank Franz-Josef (Josch) Hambsch, Belgium, for acquiring photometric observations of our target, and the anonymous referee for helpful comments and suggestions that helped to improve the paper.